# Initial growth rates of malware epidemics fail to predict their reach

**Authors:**

Lev Muchnik[1,2], Elad Yom-Tov[2], Nir Levy[3], Amir Rubin[3,4], Yoram Louzoun[5,*]

**Affiliations:**

[1]School of Business Administration, Hebrew University of Jerusalem

[2]Microsoft Research, Herzliya, Israel

[3]Microsoft Israel Research and Development Center

[4]Department of Computer Science, Ben-Gurion University of The Negev, Be'er Sheva, Israel

[5]Department of Mathematics, Bar Ilan University, Ramat Gan, Israel

[*]Please send correspondence and requests to louzouy@math.biu.ac.il

**Abstract**

Empirical studies show that epidemiological models based on an epidemic's initial spread rate often fail to predict the true scale of that epidemic. Most epidemics with a rapid early rise die out before affecting a significant fraction of the population, whereas the early pace of some pandemics is rather modest. Recent models suggest that this could be due to the heterogeneity of the target population's susceptibility.

We study a computer malware ecosystem exhibiting spread mechanisms resembling those of biological systems while offering details unavailable for human epidemics. Rather than comparing models, we directly estimate reach from a new and vastly more complete data from a parallel domain, that offers superior details and insight as concerns biological outbreaks.

We find a highly heterogeneous distribution of computer susceptibilities, with nearly all outbreaks initially over-affecting the tail of the distribution, then collapsing quickly once this tail is depleted. This mechanism restricts the correlation between an epidemic's initial growth rate and its total reach, thus preventing the majority of epidemics, including initially fast-growing outbreaks, from reaching a macroscopic fraction of the population. The few pervasive malwares distinguish themselves early on via the following key trait: they avoid infecting the tail, while preferentially targeting computers unaffected by typical malware.

**Introduction**

A key challenge of epidemiology is the prediction of the expected scale of an epidemic once initial signs of it are detected. Considering the vast number of known outbreaks in recent history, the ability to reliably predict which of them will evolve into a large-scale pandemic is essential for the deployment of efficient containment policies [1], thus limiting the epidemic's impact on the population and healthcare systems. Only a few of the approximately 3,000 local biological virus outbreaks in human populations reported by the WHO in the past 25 years [2] had evolved into global pandemics. Given the average rate of one reported outbreak every three days, the ability to predict the few that require the global mobilization of resources is critical.

Large differences between the predicted and observed scales of epidemics have been recorded time and again, leading to a recognition that projecting infectious disease outbreaks using current mean-field methods was unreliable and therefore of limited use. The failure of the existing models to cope with the challenge has been recognized by the organizations facing it. For instance, a WHO report states that "forecasting disease outbreaks is still in its infancy" [3]. Another formal report issued by the National Science and Technology Council (US) confirms that "public health response to emerging infectious disease threats has often been largely reactive – a response is mounted after an outbreak is recognized." [4]. Several initiatives (e.g. Epidemic Prediction Initiative – https://predict.cdc.gov/, FluSight – https://www.cdc.gov/flu/weekly/flusight/index.html, Dengue Fever Challenge) inspired by these and other reports were created to test the out-of-sample prediction capability of the existing models. Such attempts to face these real-world challenges have led to an acknowledgment of the failure of existing approaches to generate consistent and accurate predictions of outbreak characteristics [5-7]. For instance, a multi-year contest involving sixteen teams of epidemiologists who competed to produce the most accurate forecast characteristics for seasonal dengue fever outbreaks [6] found that the forecast quality varied widely, with particularly inaccurate predictions for high-incidence seasons.

One explanation for the gap between reality and the predictions of epidemic models could be the tendency for the effective reproduction number to decrease over time. In the case of biological viruses, this decrease can stem from multiple factors, including human intervention – [8] passive vaccination of the population [9]; behavioral changes, such as social distancing [10] – or seasonality [11] arising from the degree to which a virus's transmission and survival mechanisms depend on environmental factors [12]. Still, epidemics often develop at a slower rate than anticipated [13,14], having only a limited effect on the population (e.g. recent MERS, ZIKA, several outbreaks of Ebola [14], foot-and-mouth disease [13], different variants of influenza, and many others [15,16]) and ultimately dying away, even as insufficient measures (or none) are taken to contain them. Thus, even if the measures deployed contribute to limiting the spread, other mechanisms are necessarily involved.

Recent publications suggest that the failure of existing diffusion models to extrapolate the behavior of future epidemics from their early growth rates could be related to these models' failure to take extreme population heterogeneity into account [17-19]. The general idea is that pathogens insufficiently contagious to survive and spread in the general population may still spread at a very high rate in a heterogeneous population, specifically among exceptionally susceptible targets. However, such outbreaks cannot reach their expected spread numbers (as predicted by their rapid initial growth rate,) since the few outliers (the exceptionally susceptible) are quickly removed from the population.

To understand how a target population's heterogeneity affects its susceptibility (while parsing the implications of heterogeneity for outbreak dynamics), one would need to have detailed exposure and infection-history data for every individual in a study. Although multiple proxies for such heterogeneity have been developed [20], we currently command no large-scale data that could confirm the underlying mechanism in a great enough number of epidemics and thus support a general claim.

In a world suffused with technology and running on software, computer malware – the bane of individuals and businesses alike – presents similarities to pathogens in biological populations, both human and animal. Software security teams face challenges analogous to those faced by healthcare systems; early identification of the malware likely to affect a large number of computers is essential to the efficient management of a computer ecosystem. At the same time, in contrast to the situation with biological pathogens, there exist detailed data for a very broad number of malwares, allowing empirical analysis adapted to confirm that extreme heterogeneity of the susceptible population is indeed the key factor affecting the spread of the contagious pathogen.

Differences do exist between the propagation mechanisms of biological viruses and computer malware, including varying spread mechanisms and, naturally, the clearance mechanism, i.e. individual healing in biological systems, in contrast with malwares, which are often cleared through a central antivirus update. Still, the two spheres share many common aspects [21]. A key advantage of using malware as an analog lies in the detail level of the telemetry reports generated by anti-malware software. Recording infection spread at the machine level in granular detail, these reports allow researchers to reconstruct the history of infections for each machine and characterize in tandem each piece of malware and each infected machine.

**Results**

To understand why an epidemic's reach may be limited even as it boasts a rapid initial growth rate, we studied anonymized, machine-level data from telemetry reports for the Microsoft Defender Antivirus software (Supp. Methods). Integrated into Microsoft Windows and operating on over half a billion computers, which comprise over 50% of the machines running the Windows operating system, Defender Antivirus monitors the hard drive for malware and produces telemetry reports enabling the tracing of malware propagation. Each report contains a unique machine identifier, the infection time and a file fingerprint labeling the malware. Unlike aggregated data used by most epidemiological studies to track population dynamics over time, these details show which machines were infected by which malware, even retrospectively (i.e. machines will identify the time of infection even if a file is only identified as malware later on). These data enable the reconstruction of each machine's infection history alongside with the propagation details of every malware. Different malwares utilize different spread mechanisms, including e-mail, web, or direct file transfer. Here, we do not focus on the details of the malware spread mechanisms, but rather treat all malware spreading among computers as an ecosystem exhibiting certain statistical characteristics. We studied nearly 30M infections observed in the first 72 hours of a spread of 139,962 malware strains detected over twenty-one full months between April 1, 2017 and December 31, 2018, as well as the malware's final reach. These data cover malware affecting over 200 machines.

As demonstrated below, the discrepancy between early predictions and eventual reach may be due to the extreme heterogeneity of the susceptible population's infection probabilities. Computer susceptibility, defined as the number of malwares infecting each machine, is indeed very broadly distributed (Fig 1A). The vast majority of the machines reported not a single malware infection during the entire observation period. Among machines infected at least once, the susceptibility distribution is scale-free, with a long tail of a few computers infected by a very large number of malwares, and most computers infected very rarely (Fig. 1A). To illustrate the heterogeneity: 10% of all infections are reported by as few as 0.6% of the most frequently infected machines.

Such heterogeneity explains both the rapid initial rise in infection and the failure of predictions to effectively extrapolate subsequent epidemic dynamics. In particular, the initially infected population resides on the right-hand side of the susceptibility distribution. This tiny fraction of the total population is rapidly removed, slashing the average susceptibility of the remaining susceptible population [22,23] and leading to the premature collapse of the epidemic. In such populations, the rapid initial growth would be followed by a sharp decline in the reproduction number and the epidemic's fast decay, resulting in discordance between the expected reach – based on the initial growth – and the observed reach.

We directly tested the relationship between the outbreak's initial growth rate and its reach. Figure 1B demonstrates each malware's mean hourly growth rate, averaged for malwares of different sizes. The graph is bimodal. Its left side demonstrates that the initial growth rate correlates negatively with the final reach recorded for the malware – reaching up to a few thousand machines – in clear contrast with SIR and SIR-like models in homogenous populations. Note that this comprises the majority of all observed malwares (see the malware outbreak size distribution, Fig. 1C). In contrast, for the few malwares with the largest reach, the correlation is positive, hinting at the presence of different spread dynamics for the few very far-reaching malwares. To further demonstrate the lack of correlation between initial and later growth rates, we computed whether malware attaining at least half of its total reach within 72 hours of the first occurrence (designated as "fast") had a faster initial infection rate than malwares attaining less than half of their reach within the first 72 hours ("slow"). Figures 1D and S1 show distributions in each group of the average hourly growth rates computed over the first 12 hours of the spread of each malware. Consistent with Fig 1B, malwares with both low and high initial infection rates reach their peak rapidly, while malwares reaching their peak late are over-represented among malwares with mid-range initial growth rates.

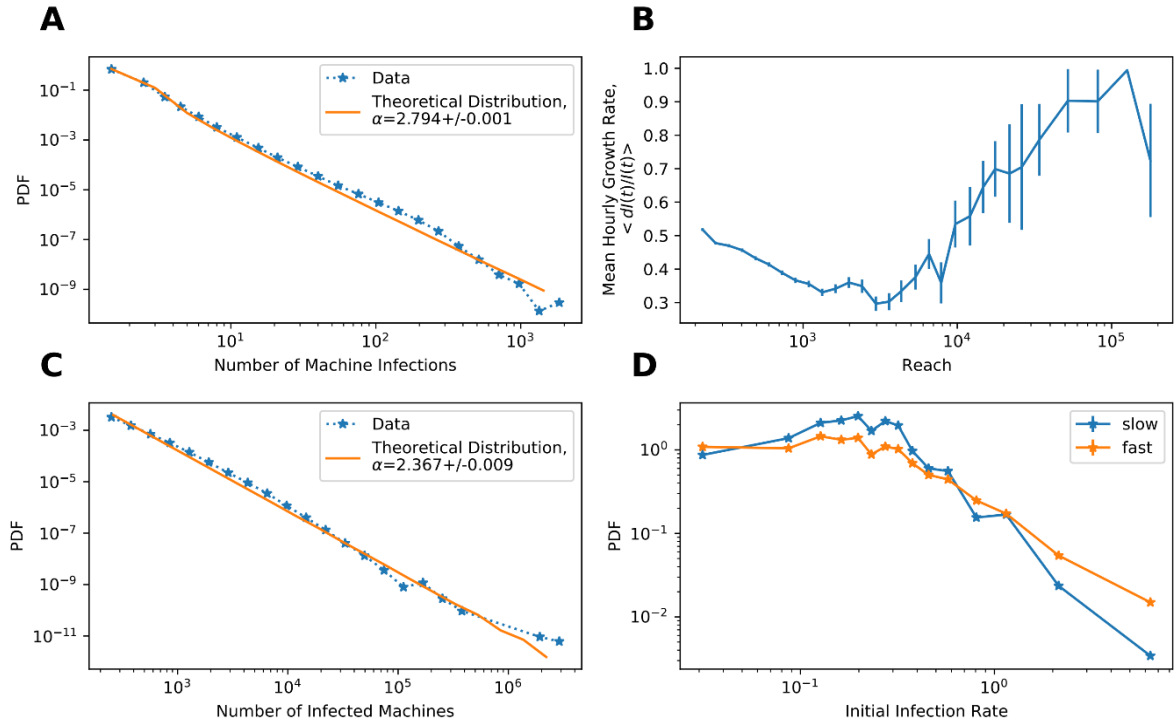

*Figure 1 A)* Probability density function (PDF) of machine susceptibility. The X-axis represents the number of distinct malwares found on a specific machine during the observation period. The distribution fits a power law with α = 2.8 – a solid line. (see Supp. Mat., Methods 1 for details). *B)* Average initial hourly growth rate computed over the first 72h of the spread as a function of the final malware reach. (see Supp. Mat., Methods 2) *C)* Malware reach PDF. The graph represents 139,920 malwares, each with a final reach exceeding 200 machines. The average malware reach is 914.6 machines, the median is 407, standard deviation is 10767. (Solid line – power-law with α = 2.367) *D)* PDF of the mean hourly malware growth rate computed over the first 12h of the outbreak for two populations: slow (under 50% of their reach in 72h) and fast (over 50% of their final reach achieved in 72h) spread.

These results may be understood from the perspective of a random mixing infection model with a distribution of susceptibility as in Fig. 1A [17], or of network infection models with a similar degree distribution. For the random mixing model, we ran stochastic SIR simulations with each susceptible node having a constant yet different probability of being infected as well as a constant probability of infecting others (See Supp. Mat., Methods 3 for details, and Melka et all for efficient modelling platform[24]). Varying the details of the simulations did not impact the results qualitatively.

The results of this modeling exercise demonstrate that (Fig. 2A):

a) The susceptibility of the infected population is higher than the average susceptibility of the population at large.

b) The average susceptibility of the infected population decreases over time, as the right-hand side of the distribution is depleted, causing a rapid decrease in the number of infected computers, which is not correlated with the initial growth rate.

These findings imply that the higher the heterogeneity of a population, the larger the susceptibility gap between those individuals infected early on and those infected later. In extreme cases, epidemics that might start spreading quickly will stop spreading once the few exceptional individuals are infected and removed. The opposite example would be the spread of infection in a homogeneous population. In that case, any outbreak will continue as long as the density of the remaining susceptible individuals is high enough to sustain the malware's reproduction (i.e. until the herd-immunity threshold is attained). In homogenous populations, a similarly skewed distribution of susceptibility could be obtained via a scale-free distribution of the probability of being exposed to the threat. Such systems are typically modeled as scale-free networks [25] in which certain highly connected nodes (sometimes designated as hubs) are in contact with a macroscopically large fraction of the population. Variation of the number of contacts that can lead to disease transfer has been shown to affect network processes, including the dynamics of disease spread [26,27]. Additionally, network hubs are known to become infected early on in the diffusion process [28]. We demonstrate this by executing an SIR simulation in artificially generated networks with power-law degree distribution, with $\alpha = 2.5$. Uneven exposure to infected individuals results in fast decay in the degree of newly infected nodes (Fig 2B).

We have tested these predictions against the observed malware spread. Indeed, the decrease in malware infectivity over time empirically illustrates predictions made by SIR simulations (Fig. 2C and Supp. Mat. 4). This phenomenon results in an overestimation of the epidemic's reproduction number and its reach from inception on. The initially observed infectivity is not characteristic of the entire population, rather decreasing gradually as the epidemic grows.

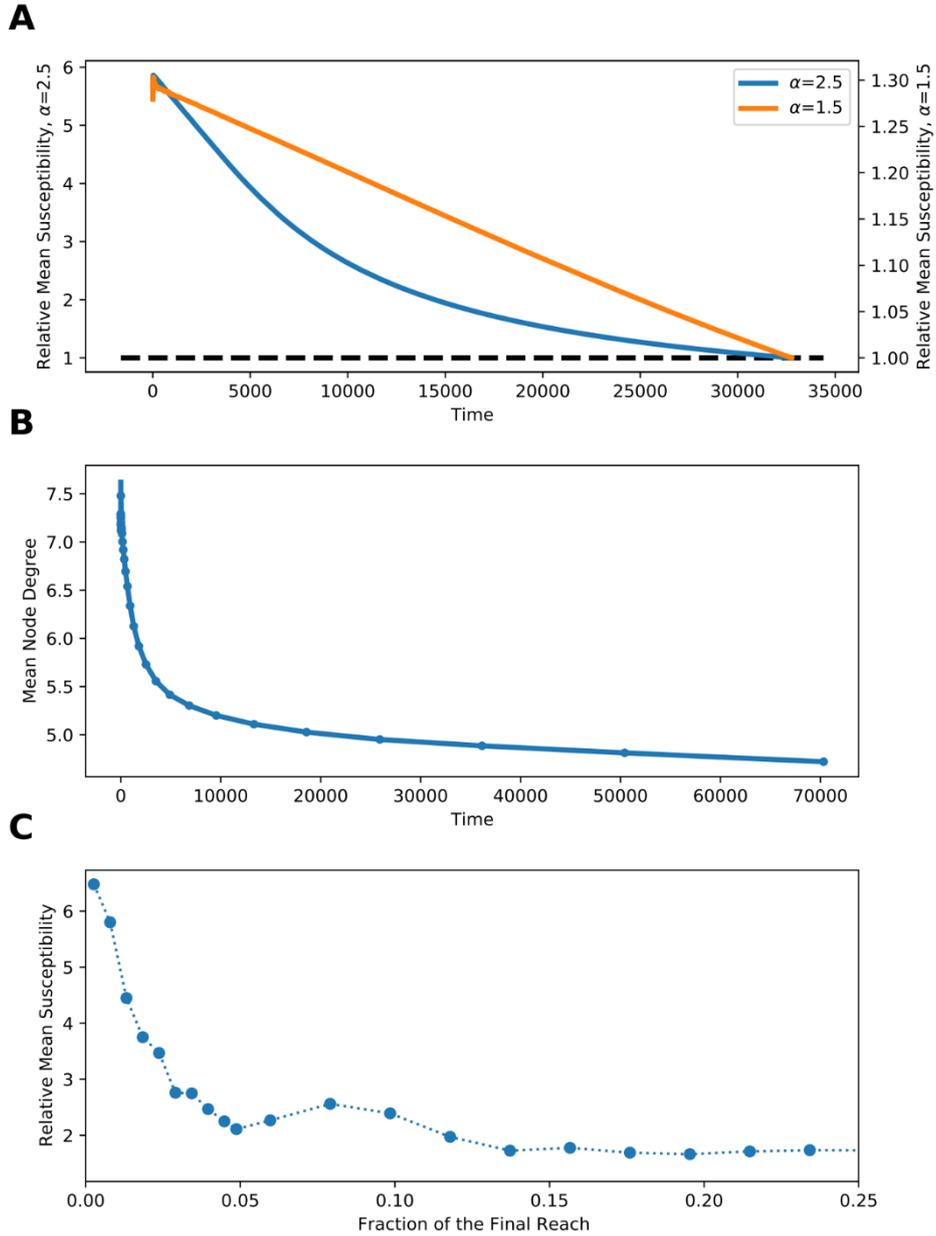

*Figure 2* Mean susceptibility of the infected individuals for the SIR simulation run for **A)** scale-free distribution of susceptibility with $\alpha = 2.5$ and $\alpha = 1.5$. Dashed line represents mean population susceptibility and **B)** The average degree of infected individuals in a scale-free network with homogenous susceptibility and power-law degree distribution, with $\alpha = 2.5$. The network was created using the configuration model. Panel **C)** shows the ratio of the susceptibility of newly infected machines for each malware as it spreads, to the susceptibility of the machines infected by the same malware, with their infection times shuffled. This plot shows the average ratio over all malwares that reach 85% of their final spread in the first 72h.

And yet, scale-free networks that facilitate the spread of infections cannot explain the dynamics observed in the real world. Per the mechanism detailed above, one would expect all epidemics to die out before affecting a large fraction of the population. All epidemics spreading on scale-free networks are destined to start in the tail of the highly connected nodes and terminate as they reach the bulk of the less susceptible population. At the same time, the data reveal that some malwares do reach a large fraction of certain populations. Such pandemics may have a different spread mechanism. A clear differentiating factor between high-reach computer pandemics and the vast majority of malware epidemics is the low average infectivity of the former (Fig. 3A) along with their high initial growth rate (Fig. 1B). These two phenomena suggest that large pandemics do not get large merely on the back of their high infectivity. In fact, they do so primarily by avoiding typically infected machines and excessively spreading among the populations usually left unaffected. By selectively targeting the main body of the low-susceptibility population, such an agent gains a much wider reach than others.

We verified this hypothesis by analyzing the properties of the computers infected by epidemics of different scales. While all malware tends to infect more low-susceptibility than high-susceptibility computers (i.e. since the vast majority of computers have a very low susceptibility, Fig. 1A) in the first 72 hours, the balance in high-reach malware is shifted toward low-susceptibility computers.

This is clear from the susceptibility distributions of the machines infected by malwares, ranked by their reach (Fig. 3B). The distribution steepness drops with malware rank, revealing that the share of frequently infected machines in the large outbreaks drops significantly. The plots in Fig. 3B are all ordered by their total reach, demonstrating that larger malware outbreaks correspond to the distributions with a very small share of highly susceptible machines. Essentially, malwares that will eventually evolve into large pandemics are characterized by the relative homogeneity (by susceptibility) of their target populations (Fig S2). To demonstrate that this is not a circular argument, we simulated the spread of each malware while preserving the susceptibility of every machine and the reach of the malware (Table S1). In this null model, all malwares have similar properties, and in each infection event a machine is selected based on its susceptibility (See Supp. Mat. for null model details). In such a simulation, the infected machine's susceptibility is not associated with the malware's reach (Fig. 3C). An important result of the initial concentration on low-susceptibility targets is the possibility of predicting a malware's reach early. In contrast with the initial slope of the epidemic's growth (Fig. 1B), the characteristic susceptibility of the affected machines is an excellent classifier of high-reach epidemics.

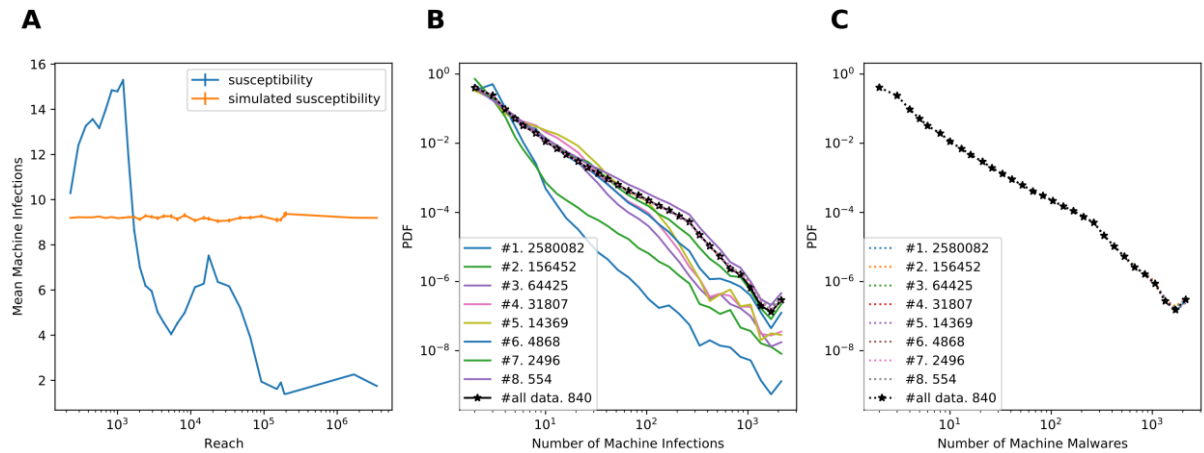

*Figure 3 A) The mean susceptibility of the machines affected by malware during the first 72h as a function of final reach. The orange line is the mean simulated susceptibility of a random equal number of machines. For each malware, the simulation "infects" computers randomly, with the probability proportional to their observed susceptibility, until the observed reach of the malware is obtained. B) Distribution of machine susceptibility for the machines affected by malware, grouped by their reach. The black line represents all malware. The legend shows the mean epidemic's reach for each group. See Table S1 for the details of group composition. C) Same distribution and grouping as in B), but these results are generated by a simulation that randomly assigns machines to malwares according to each machine's suceptibility and the malware size. The black line represents the PDF of the actual distribution of susceptibility for the entire population. As may be clearly seen in 3C, the machine's susceptibility precisely follows the PDF and is not associated with the reach.*

**Discussion**

Classical epidemiological models frequently fail to predict the reach of diseases based on the initial rise in the number of those infected. Even for the same pathogen, the reach has varied dramatically between countries and even areas of the same country [29-31]. A recent survey of the measles reproduction number [32] collected 58 reported values, with most located in the 4-18 range but demonstrating a long tail reaching a value of 770.38, set by Wallinga et al. [33]. Similarly, a large difference between districts was reported for the 2014-2015 Ebola outbreak in Western Africa [13]. The lack of correlation between the epidemics' early spread rates and their later dynamics was systematically confirmed by several out-of-sample studies and has been a source of concern for the relevant authorities. This variability in the epidemics' spread dynamics may be due to the extreme sensitivity of the contagion to the volume of the susceptibility-distribution tail [17-19].

While the specific results we have shown stem from an analysis of malware spread, they are a direct consequence and a characteristic example of heterogeneity and are therefore widely applicable. The conclusions may be extensible to other contact processes, such as the spread of products, pathogen-driven epidemics, and information cascades [34]. It should be noted that our analysis became possible due to the availability of exhaustive historical records on an individual machine's propensity to become infected by the spreading agent. Such data are rarely available for biological pathogens. Several studies performed meticulous contact tracing to find extreme individual heterogeneity of infectivity and relate this heterogeneity to a variation of the disease reproduction number [20], confirming dynamics similar to the one we report for malware. At the same time, even if a significant effort were to be invested in monitoring epidemics, much of the necessary information would still go unobserved. For instance, less than 10% of the estimated fatalities during the 2009-2010 influenza H1N1 pandemic were laboratory-confirmed [35]. The fraction of confirmed COVID-19 patients is still debated and could vary significantly by region, but is estimated by
 some studies as hovering between 4 and 14 percent of all cases [36,37], suggesting that the true dynamics of the disease spread are seldom observed. Furthermore, even when the infection status of most patients is known, the detailed medical records necessary to assess the patients' characteristic susceptibility to infectious diseases are rarely available. The reconstruction of machine-level infection history for the entire ecosystem (feasible through the availability of Microsoft Defender telemetry reports) greatly facilitates a comprehensive analysis of how a variety of malware interacts with the computer population. The primary insight arising from this analysis is the surprising fact that the malware with the largest reach affects the tail of the susceptibility distribution significantly less than the majority of malware strains. By spreading in a relatively homogenous population, such malware is characterized by a lower variation of the reproductive number. Its subsequent trajectory is therefore easier to predict.

The testing and application of our findings to other domains would require collecting similarly detailed data for the corresponding systems. We have not studied here what property of the malware determines the differential targeting patterns. Understanding this factor may be key to fighting epidemics with a very wide reach, or, inversely, to developing a successful product or promoting social change. These results point to an effective intervention policy. Instead of looking at the effect on R0 (or on the infection probability matrix, when the population is segregated into subgroups), one should analyze the composition of the infected population. As with strategies developed for scale-free networks, targeting the tail would be a very effective way of combating the vast majority of (small-scale) malwares [38,39,40].

Still, such a strategy is inefficient for preventing very large outbreaks. Small-scale actions targeting the highly susceptible population would slash the effective R0 and terminate the epidemic at its inception. Conversely, reducing the susceptibility of each potential target via broader intervention would prolong the epidemic, since the spread could still be sustained in the tail. To apply such models meaningfully to biological viruses, models for the susceptibility of each potential target based on its demographics and behavior must be developed, with detailed data collected to fine-tune such models.

**Acknowledgments**
This analysis was supported by the Israel Science Foundation, grant no. 1777/17.